\documentclass[11pt,a4paper]{article} 
\usepackage{caption}         %
\usepackage{charter}         
\usepackage{graphicx}        
\usepackage{amssymb,amsmath} 
\usepackage{multicol}        
\usepackage{url}             
\usepackage{enumitem}        
\usepackage{marvosym}        
\usepackage{wrapfig}         
\usepackage[T1]{fontenc}     
\usepackage{mathptmx}        
\usepackage{datetime}        
\usepackage{fancyhdr}        
\newdateformat{mydate}{\monthname[\THEMONTH] \THEYEAR}          
\usepackage[pdfpagemode=FullScreen, colorlinks=false]{hyperref} 
\usepackage{booktabs}        
\usepackage{algorithm}
\usepackage{algorithmic}
\usepackage{array}
\usepackage{hyperref}                                      %
\hypersetup{
    colorlinks = true,
    citecolor  = blue,
    linkcolor  = blue,
    urlcolor   = blue
}

\newcommand{\NewsTitle}[1]
{
    \begin{center}
    \usefont{T1}{ptm}{b}{n}               
    \vspace{14pt}\Large #1\vspace{1pt}    
    \par \normalsize \normalfont
    \end{center}
}

\newcommand{\NewsAuthor}[1]
{
    \begin{center}
    \usefont{T1}{ptm}{n}{n}               
    \textsc{#1} \vspace{4pt} 
    \par \normalfont
    \end{center}
}

\graphicspath{{./figs/}}
\usepackage{subfigure}

\usepackage{amsmath}
\usepackage{graphicx}
\usepackage{multicol}
\usepackage{textcomp}
\usepackage{booktabs}
\usepackage{enumitem}

\usepackage{algorithm}
\usepackage{algorithmic}
\usepackage{balance}

\begin{document}



\setcounter{section}{0}
\setcounter{figure}{0}
\graphicspath{{./figs/}{./figs/item-yiyu/}}
\NewsTitle{Real-Time Boiler Control Optimization with Machine Learning}
\NewsAuthor{Yukun Ding, Yiyu Shi\\ University of Notre Dame}

\section{Introduction}
As coal-fired power plants currently produce 
41\% of global electricity \cite{worldcoal}, proper control of coal-fired boilers in producing
electricity is not only essential to the safety of power plant operation, but also directly affects
boilers stability, energy efficiency, and sustainability, thus having huge socioeconomic and environmental impacts \cite{hasini2009analysis}.
How to optimally control boilers' operating condition in real-time is, however, difficult. The combustion process inside a boiler is highly complex and nonlinear with strong-coupling and time-delayed influences. It is well understood in literature
that it is not easy to achieve high efficiency in operating large utility boilers and most existing practices in the industry are highly sub-optimal \cite{feng2009combustion}. 

Nonuniform temperature distribution inside a boiler is known to cause tube rupture, a frequent failure mechanism for boiler operations. But to maintain a uniform temperature distribution inside a boiler is difficult even for domain expert in practice due to the dynamic air flow inside the boiler.
One of the most frequently used practices to deal with nonuniform temperature 
distribution is spraying water inside a boiler, which introduces unnecessary efficiency loss and additional operating cost \cite{hasini2009analysis,park2013numerical}.
Another practice is to remold a boiler by re-arranging super-heater panels to alleviate the uneven temperature distribution \cite{yin2003further}, which requires to shut down the boiler and cannot be done in real-time.
Temperature distribution inside a boiler has been studied using various computational fluid dynamics (CFD) methods under steady-state conditions \cite{hasini2009analysis,park2013numerical}. However, employing the CFD methods for real-time boiler control is not feasible because of the extremely high computational cost of solving the CFD models \cite{diez2005modelling}.

To avoid solving the physical-based CFD models, researchers have proposed to use machine learning-based
methods to predict boiler temperature or other related parameters in order to control boiler combustion process \cite{li2017deep,li2016combustion,zhao2009modeling}. However, such formulations have mainly focused on reduction of pollutant emission, not for the uniform
distribution of temperature inside boilers. 
There are some works focusing on the boiler efficiency optimization \cite{gu2011online,kusiak2006combustion} where external measurements (e.g. exhaust gas temperature) are used to estimate the boiler efficiency through some models due to the difficulties in obtaining temperatures within the boiler. Instead, in this work we have collected the temperature distribution data within the boiler from our industry partner, which allows precise and accurate observation of the combustion efficiency and stability.
Moreover, due to the strong-coupling, nonlinear and large time delay characteristics of  boilers,  existing solutions using neural networks often result
in a complicated black-box optimization problem and thus can hardly ensure good real-time performance \cite{li2016combustion,xu2013balanced,zhao2009modeling,feng2009combustion}. 

In this paper, we use a new formulation to the boiler control optimization problem based on inputs from industry, i.e., maintaining a uniform distribution of temperature in different zones and a balanced oxygen (O$_2$) content from the flue in a coal-fired power plant. We develop a new but practical solution framework to solve the proposed real-time control optimization problem by combining machine learning and optimization techniques. We validate the formulation and show high solution quality using a real industry boiler dataset.  Our results suggest that, in specific scenarios, a dedicated system with simpler models can be more desirable than using more powerful models in terms of both performance and computational efficiency.

The rest of the paper is organized as follows. We first give a background about the boiler control
problem. Then we present our formulation of the problem and the solution. We
then report the experimental results and make conclusions.

\section{Background}
\label{sec_back}

We give a brief introduction of the operation of a power plant boiler.  
Pulverized coal is fed into the furnace from different coal feeders with a proper volume of airflow, both of which are
controlled by their respective throttle openings to maintain a desired air-to-fuel ratio for combustion. The water circulates in a water-steam system and
absorbs the radiation energy from the furnace continuously until it  becomes high-pressure superheated steam in the superheater. Through the steam turbines, the thermal energy is transferred to mechanical work and finally becomes electricity through generators.
In a power plant, a central controller determines the desired setpoints for various subsystem controllers, a critical one of which is the combustion controller that determines the feed rates of coal and airflows \cite{liu2010nonlinear}.

Since combustion quality ultimately determines the production efficiency, we focus on combustion control in this paper. 
In general, higher temperature inside the furnace and lower O$_2$ content in the flue indicate higher efficiency. But
to ensure sustainably high combustion efficiency and stability, it is desirable to also maintain a 
balanced high temperature distribution and low O$_2$ content in the flue, as a balanced distribution of both temperature and O$_2$
content indicates that both flames and pressures are uniformly distributed, and thus promising the stability and safety of the boiler.
However, existing formulations \cite{li2016combustion,xu2013balanced,kusiak2005optimizing,zhao2009modeling} have not considered the
temperature distribution, and not mentioned the distribution of O$_2$ content.

In current industry practice, the combustion controller consists of a set of PI/PID controllers and pre-computed set-pints (computed theoretically and fine-tuned empirically). The well-established control system based on PI/PID was improved by advanced PI/PID controller such as auto-tuning PID \cite{zhang2012simultaneous}. In recent years, machine learning-based prediction control has been studied widely and included in commercial boiler control solutions \cite{NeuCo,kusiak2006combustion}. The prediction model was used for steady state optimization initially and then gradually for real-time optimization \cite{pechenizkiy2010online,szega2015optimization,liu2010nonlinear,foreman2008architecture}.
From the algorithm perspective, the modeling approaches are dominated by neural networks and their variants, e.g. vanilla feed-forward network, radial basis function (RBF) network, double linear fast learning network \cite{li2017deep,li2016combustion,feng2009combustion}. The optimization problems are solved by various heuristic search algorithms, e.g. genetic algorithm (GA), differential
evolution (DE), particle swarm optimization (PSO), ant colony optimization (ACO) \cite{zheng2008combining,peng2009improved,zhang2007modeling}. Even though the recent advancement on more computationally efficient neural network \cite{ding2018universal,xu2018scaling,xu2018quantization,liu2018pbgan,rastegari2016xnor,xu2018efficient,han2015deep,xu2018resource}, due to the considerable number of variables, the computational requirement remains a challenge which degrades the real-time performance \cite{zhao2015fuzzy}.

\section{Problem Formulation and Solution Framework}
We formulate a new boiler control problem in this section by not only maintaining a high temperature and low
O$_2$ content, but also maintaining a uniform distribution of temperature in different zones and 
O$_2$ content inside the flue in a coal-fired power plant. The goal of achieving a balanced distribution of temperatures and O$_2$ content can be captured by a quadratic penalty function of the deviation of temperatures from the average value and the difference between O$_2$ content from two sides in the flue. Certainly there are other options but quadratic penalty function is employed, because it is differentiable, suitable for capturing the deviation from the desired value, and relatively simple for optimization.
For effective combustion, we also want to maintain a high temperature and low O$_2$ content inside a boiler.
Together, we can use a weighted sum of these components as our objective with the constraints under real operation such as the given range of controllable variables and their sum. The polynomial objective function has four terms, indicating the variance of temperature in different zones, the difference of O$_2$ content in two sides of flue, 
the average temperature, and the average O$_2$ content, respectively. The problem needs to be solved continuously for every time stamp $t$ based on data including operations from $t-1$ and prior
in order to achieve the goal of real-time control for the boiler. $f_i^T$ and $f_j^O$ define prediction models for temperature and O$_2$ respectively where $i$ and $j$ denote the index of models for temperature and O$_2$ content at different zones.

\begin{figure}[ht]
  \centering
  \includegraphics[width=3.2in]{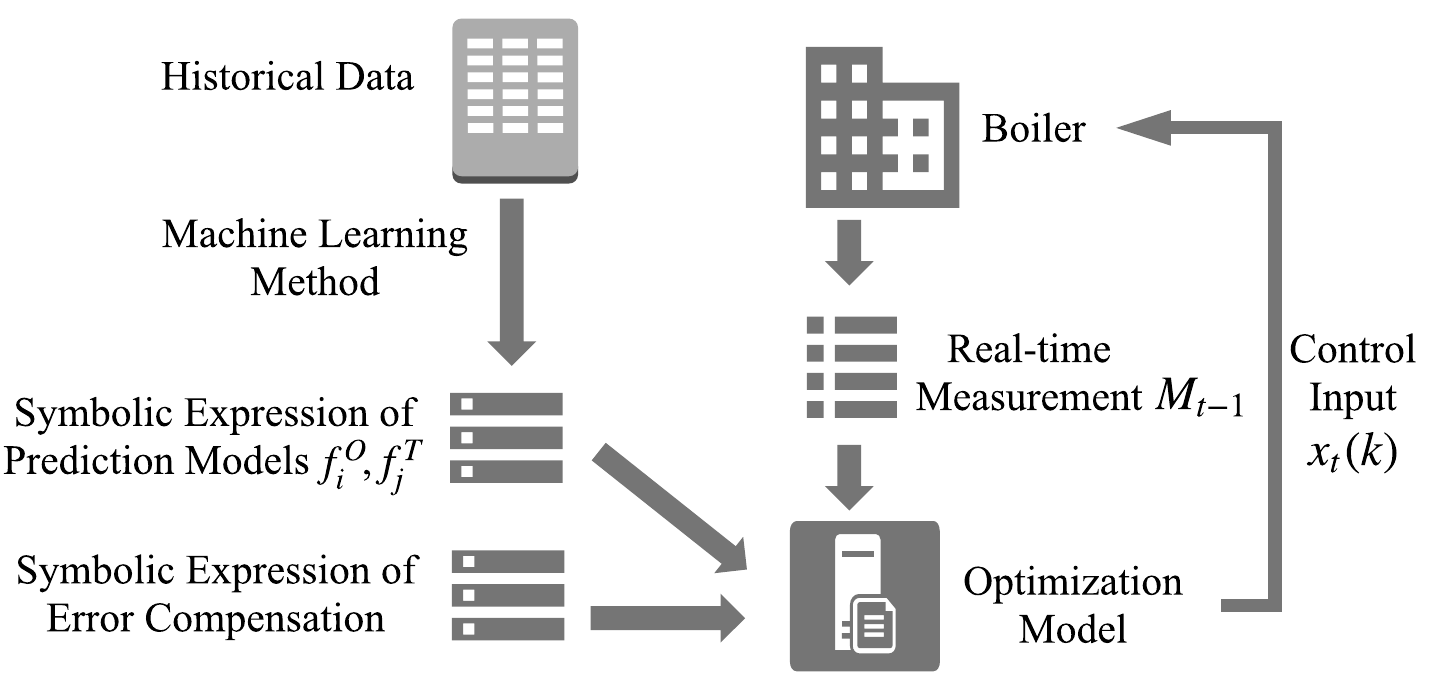}
  \caption{Structure of the solution framework}
  \label{fig:structure}
\end{figure}

The structure of proposed real-time boiler control framework is shown in Figure~\ref{fig:structure}.
The prediction models, $f^T_i$ and $f^O_j$ trained on historical data, provide the symbolic expression of temperature and O$_2$ content based on control variables and other measured uncontrollable variables, which are denoted as $x_{t-1}$ and $M_{t-1}$ respectively. In every time step, $M_{t-1}$ in the symbolic expression will be replaced by the latest observed values and only the controllable variables $x_{t-1}$ and the optimization objective $V_t$ remain. Then the optimization model takes the resulted expression and solves the nonlinear programming problem to give the optimal combination of controllable variables, which is the control input to the boiler. An error compensation module, which will be discussed later, is employed to further improve the prediction accuracy. The time cost for solving the optimization model at every time step depends on the choice of optimization algorithm and the complexity of the problem determined by the prediction model.
Since the control loop needs to be continuously solved for every time step as soon as possible, the runtime performance of the optimization model is the critical consideration.

We employ machine learning-based approaches for predicting both temperature and O$_2$ content. 
We notice that there is a special mathematical structure of the given problem that the constraints are linear with respect to controllable variables $x$ 
and the objective is quadratic with respect to the predicted values $T(i)$ and $O(i)$. 
Therefore, among many possible choices of machine learning techniques, we use the
epsilon-support vector regression ($\varepsilon$-SVR)  with linear kernel \cite{vapnik2013nature}
as the prediction model.
Such a choice will render a nice mathematical 
structure for the optimization model, which in turn enables us to choose an effective
optimization technique to solve the problem efficiently. We obtain the linear prediction models
through the $\varepsilon$-SVR linear kernel method. Plugging the function of the prediction models back into the objective function with some rearrangement, we obtain a quadratic
programming model as follows (by dropping the subscript $t$ for simplicity):
\begin{equation}
\label{equ:qp}
\begin{aligned}
    \underset{x}{\rm{min}}&\quad V(x)=\frac{1}{2}x^\mathrm{T}Hx+f^\mathrm{T}x+c\\
    s.t &\quad  A_{q}\cdot{x}\le b_{q};
        \quad  A_{e}\cdot{x}= b_{e};
        \quad  B_l\le x\le B_u.
\end{aligned}
\end{equation}
\noindent where $H$ is a real symmetric matrix of coefficients, $f$ is a vector of coefficients, and $c$ is a constant term, all of which 
can be easily constructed based on values from the prediction model.
$A_q$, $A_e$, $b_q$, $b_e$, $B_l$, and $B_u$ are compact representation of known constraints. 

Therefore, at each time step, we end up with a quadratic programming problem for the optimization model. We adopt an efficient algorithm for quadratic programming, the interior point convex (IPC) method 
\cite{gould2004preprocessing,mezura2006comparative,gondzio1996multiple}.
It uses a presolve procedure 
to remove redundancies and simplify constraints. It then tries to find a point where the Karush-Kuhn-Tucker (KKT) 
conditions hold, and use multiple corrections to improve the centrality of the current iteration.

As discussed before, the working mechanism of a coal-fired boiler is
extremely complex and time-varying, and not all factors are observable through
measurements. Therefore, a machine learning-based prediction model of this kind
may produce time-related local bias because of the change of underlying hidden factors,
such as the fluctuation of coal quality and the restart of boilers. 
By borrowing an idea from the field of control,
we add an error compensation part to further improve the prediction accuracy by compensating the local bias, which is estimated by computing an average difference 
between the actual output and predicted output for a prior window size of time steps, and
adding this value to the future predicted output to decrease residuals \cite{zhang1985error,draper1966applied}. At every time step, the latest prediction error is obtained and the new compensation value is added to $T_{t}(i)$ and $O_{t}(j)$ as constants. The window size $S$ is another
algorithmic tuning parameter of this method. Note that the error compensation can be effective because the input and output are sampled from a physically continuous system, thus the adjacent prediction errors may implicitly stores contextual information and can be used for better prediction. This work is also an example of how a well-trained machine learning model can be improved by leveraging its physical meaning. This approach can be extended to other applications, where prediction is made about a continuous system and prediction error is available in online operation.

 The prediction model is trained offline and does not need to be re-trained any more. The time to solve the optimization problem dominates the latency in the control loop which is the lag from the observation to the corresponding control operations. Because the lag is not taken into consideration when building the prediction model according the dataset, the closer the lag is to zero the better. It is also the reason to simplify the complex highly nonlinear optimization problem to a quadratic programming problem, which is very important for a real-time solution. 

\section{Experimental Results}

\subsection{Experiment Setup}
We conduct experiments using the real dataset collected by our industry partner 
from a production power plant boiler as discussed before. It contains more than 13,000 samples collected in a span of more than two months at a sample rate of 432 seconds. Each sample corresponds to a time stamp with 49 features including temperatures in six zones,  O$_2$ content in two sides of flue, generation load, Nitric oxide in two sides of flue, twelve
coal feed rates, and sixteen throttle openings, etc.

For comparison purpose, we also implement different algorithms to show the effectiveness
of our proposed algorithm. The alternative options used for the prediction model include
the $\varepsilon$-SVR with a RBF kernel, the
classic three layer feed forward neural network (NN) with tangent-sigmoid activation
function for the hidden layer, the vanilla recurrent neural network (RNN) \cite{zhang2014sequential}, and
the LSTM model \cite{hochreiter1997long}. 
The alternative options for the optimization model include some popular
heuristic search algorithms, including GA, DE, PSO, and Sequential Quadratic
Programming (SQP) \cite{nocedal2006sequential}. 
All tuning parameters are selected by Bayesian optimization \cite{snoek2012practical} or grid search on a validation dataset.

\subsection{Comparison of Prediction Models}
We first compare the prediction accuracy among the five prediction models. For each model,
there are also different ways of organizing the input data (or feature selection
for $\varepsilon$-SVR based methods). Three variants are considered:
(A) non predicting data from the current time stamp, (B) all data from the
current time stamp, and (C) all data from both the current time stamp and
a varying number of past time steps. 
Most existing work on boiler optimization uses the type (A)  data \cite{zhao2009modeling,xu2013balanced,feng2009combustion}
as they assume a steady state model. Type (B) data is a special case of type (C)
data with zero previous time step data.

To show the importance of organizing input data properly, we 
apply all the three types of data to the first three methods,
and only type (B) data to RNN and LSTM models. The reason for the latter
is that RNN and LSTM needs time-dependent data and
the models themselves can be trained to capture the time-delayed effect
through internal memories.
The accuracy metrics used are averaged Mean Squared Error (MSE) and mean absolute percentage error (MAPE) of the six zones for temperature and two side for O$_2$. The prediction accuracy for temperature is reported in Table~\ref{tab:CompareModelT}.
As it can be seen, for the first three methods, results from type (B) and (C) data
are significantly better than those from type (A) data and results from type (C) data are the best. 

\begin{table}[h]
  \caption{Temperature prediction models}
  \label{tab:CompareModelT}
  \centering
  \begin{tabular}{cccccccc}
    \toprule
    {Model} & {Data Type} & {MSE}&{MAPE}\\
    \midrule
            & (A) & 975.3& 2.06\%\\ 
SVR (linear)& (B) & 289.2& 1.12\%\\ 
            & (C) & \textbf{164.8}& \textbf{0.82\%}\\ 
\hline
          & (A) & 1860.1& 2.88\%\\ 
SVR (RBF) & (B) &  1199.8& 2.24\%\\ 
          & (C) & 246.6& 1.01\%\\ 
\hline
   &  (A) & 1268.8& 2.55\%\\ 
NN &  (B) & 344.4& 1.23\%\\ 
   &  (C) & 181.0& 0.87\%\\ 
\hline
RNN &  (B)& 635.6& 1.89\%\\ 
LSTM &  (B)& 841.7& 1.98\%\\ 
    \bottomrule
  \end{tabular}
\end{table}

\begin{figure*}
  \centering
  \includegraphics[width=0.95\textwidth]{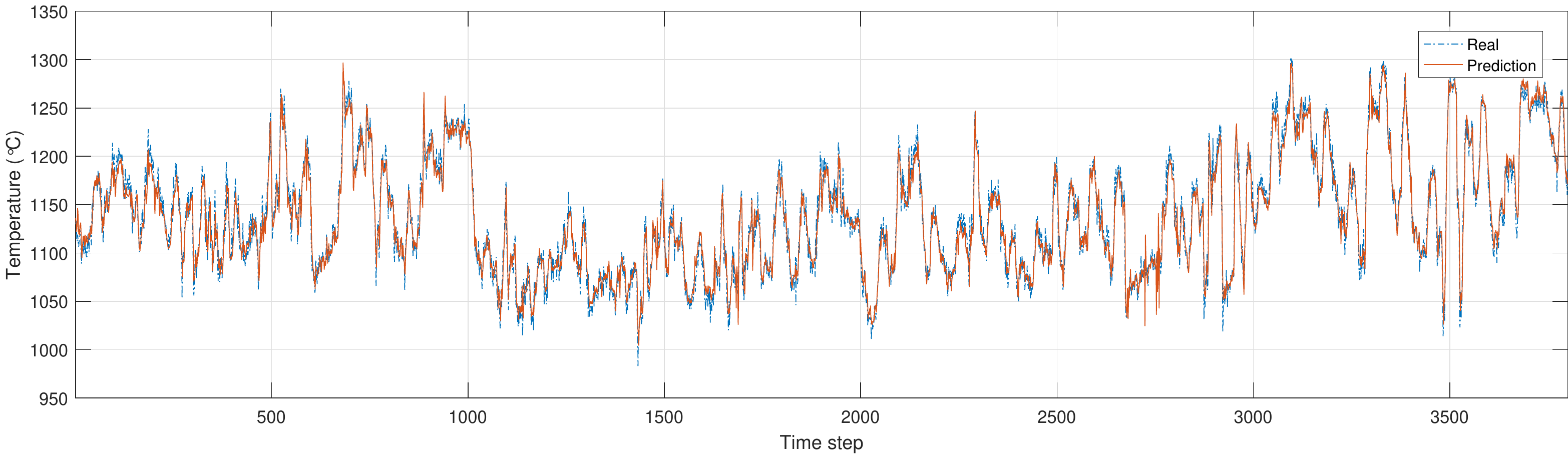}
  \captionof{figure}{Comparison of real temperature and predicted temperature in zone 1}
  \label{fig:PredictionT1}
\end{figure*}

In terms of methods, although RNN and LSTM seem to be the most suitable models for time series data, 
we suspect the limited data size (albeit one of the largest in the literature) and the peculiarity of the system dynamics have prevented RNN and LSTM from achieving a stable solution such that
the hidden states memorizing  past information are weaker than raw data from past steps when supporting the subsequent predictions. The proposed $\varepsilon$-SVR linear model performs the
best with the least average MSE value.
Even when comparing results from type (B) data for
all five methods, $\varepsilon$-SVR linear is still better than
RNN and LSTM. The temperature prediction result on test dataset by $\varepsilon$-SVR linear model is illustrated in Figure~\ref{fig:PredictionT1}.

We also offer the following reasons to explain why our proposed $\varepsilon$-SVR with linear kernel performs the best. 
First, the measurements of temperature and O$_2$ content contain some outliers because of 
the unstable airflow inside the boiler. The $\varepsilon$-SVR with $\ell_1$ loss is less
sensitive to such outliers when compared to the $\ell_2$ loss in other methods. 
Second, $\varepsilon$-SVR treats errors less than $\varepsilon$ as zero, and 
is thus less sensitive to sensor noises than others, which further
helps to reduce unnecessary updates in the training process. Third,
the simple linear regression is less likely to be overfitting 
compared to those more complex nonlinear models. Moreover, even though NN can provide better prediction performance, it cannot be used in the control system as it lead to a highly nonlinear optimization problem which is too complex to be solved in real-time. The results of O$_2$ prediction is quite similar to that in the temperature prediction.

\subsection{Impact of Error Compensation}
\begin{figure}
  \centering
  \includegraphics[width= 0.4\textwidth]{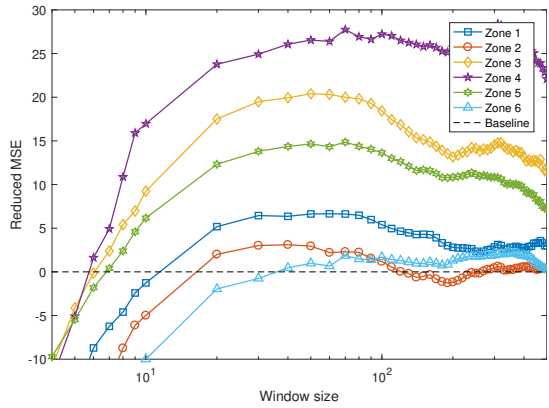}
  \captionof{figure}{The reduced MSE versus window size of error compensation} 
  \label{fig:ErrorComp}
\end{figure}

In this section, we report the impact of error compensation on the solution quality. Different window sizes can be tried on the historical data and the window size $S$ can be selected with a trade-off between complexity and performance.
Figure~\ref{fig:ErrorComp} shows the impact of different window sizes
of error compensation on temperature prediction in different zones. The window size in x-axis indicates how many previous samples are used to calculate the error compensation, which is the average error for previous predictions. The y-axis stands for how much MSE are reduced by using error compensation with a given window size and thus the higher the better.

A rough rise and fall trend can be observed as expected in Figure\ref{fig:ErrorComp}. When the window size is small, which means only a few latest errors are used to calculate the compensation, the error compensation makes prediction worse (negative values in Figure\ref{fig:ErrorComp}) because high randomness dominates the compensation. When window size increases, the compensation helps to reduce prediction errors and these curves reach a peak since a proper window size enables us to discover a local bias covered by randomness. As the window size keeps increasing, these curves fall down as too many prediction errors from long ago are used. If the window size reaches a very large value, all curves will finally converge to a narrow range around zero 
because it leads compensation to a near-zero value, which is not shown within this figure. It is worth noting that predicted temperatures with lower accuracy tend to get more improvement from error compensation. We suspect those zones are more sensitive to some hidden factors and thus have more apparent local bias. Similar observations also hold for O$_2$ content prediction. We finally select 50 as the window size for error compensation calculation to strike a right balance. 
With error compensation, we further reduced the average MSE of temperatures and O$_2$ content by 7.4\% and 3.4\% respectively.

More complicated approaches such as SVR and NN also have been tested, but surprisingly, even though they get better result under some settings, the simplest average strategy gets the best and most stable performance overall under this practical circumstance. This is probably caused by the high randomness of the boiler system and its variance along with time.

\subsection{Comparison of Optimization Algorithms}
We compare various optimization algorithms on the test dataset using the best prediction model obtained in last section. At each time step, the optimization algorithm will produce
one objective value and
for all test samples, the objective values are collected for each model. We report the comparison results in Table~\ref{tlb:CompareOptimization}, where
the solution quality is measured by the objectives collected for all
test samples, and we report their mean, minimum, maximum and standard
deviation value for simplicity. The smaller the objective value, the better the solution quality. The computation time is measured by the time to converge in
seconds on a desktop with an Intel i5-4590 3.3GHz CPU. 

\begin{table}[h]
  \caption{Comparison of different optimization algorithms}
  \label{tlb:CompareOptimization}
  \centering
  \begin{tabular}{cccccccc}
    \toprule
    {Solving} & {Time} & \multicolumn{4}{c}{Objectives}\\
    \cmidrule{3-6}
    { Algorithm } & { (sec) } & Mean & Min & Max & Std  \\
    \midrule
IPC& \textbf{0.16}& \textbf{0.085}& \textbf{-0.207}& \textbf{0.419}& \textbf{0.140}\\
DE & 81.5& 0.127& -0.168& 0.497& 0.145\\
SQP& 159& 0.117& -0.189& 0.434& 0.138\\
PSO & N/C& 0.235& -0.121& 0.599& 0.151\\ 
GA & N/C& 0.586& 0.158& 1.093& 0.234\\ 
    \bottomrule
  \end{tabular}
\end{table}

We see from Table~\ref{tlb:CompareOptimization} that only IPC, DE, and SQP can provide a converged solution within the
given time interval, while PSO and GA cannot. 
IPC outperforms DE and SQP on both runtime and result quality significantly. 
This is expected, as IPC is a most suited optimization algorithm for the
special mathematical structure as derived in this work, while other
algorithms are generic optimization techniques.

Based on the same prediction model, we observe that solutions from IPC based control
are able to reduce the temperature standard deviation by 42.5\%, and  O$_2$ content difference by 61.5\%
when compared to the the original test data without optimization. At the same time, we see 32\textcelsius{}  higher average temperature and 38.6\% lower average O$_2$ content,
indicating that the proposed model can also improve combustion efficiency simultaneously.

\section{Conclusions}
 Equipped with the unique dataset collected from a real power plant, we introduce a new formulation for boiler control problem that focuses on maintaining not only
 high temperature and low O$_2$ content, but also a balanced distribution of temperature and O$_2$ content. To overcome the foremost challenge of
 developing a real-time solution, we propose a new algorithmic framework that incorporates a machine learning-based prediction model,
 an optimization model, and an error compensation model. Experimental results validate the effectiveness and efficiency of the solution. The solution framework can be extended to other Cyber-Physical Systems where the online control or optimization is constrained by the complexity of prediction and its formulation.

\clearpage


\begin{thebibliography}{10}
\small



\bibitem{conn1997globally}
A.~Conn, N.~Gould, and P.~Toint.
\newblock A globally convergent lagrangian barrier algorithm for optimization
  with general inequality constraints and simple bounds.
\newblock {\em Mathematics of Computation of the American Mathematical
  Society}, 66(217):261--288, 1997.

\bibitem{diez2005modelling}
L.~I. D{\'\i}ez, C.~Cort{\'e}s, and A.~Campo.
\newblock Modelling of pulverized coal boilers: review and validation of
  on-line simulation techniques.
\newblock {\em Applied Thermal Engineering}, 25(10):1516--1533, 2005.

\bibitem{ding2018universal}
Y.~Ding, J.~Liu, J.~Xiong, and Y.~Shi.
\newblock On the universal approximability and complexity bounds of quantized
  relu neural networks.
\newblock {\em arXiv preprint arXiv:1802.03646}, 2018.

\bibitem{draper1966applied}
N.~R. Draper, H.~Smith, and E.~Pownell.
\newblock {\em Applied regression analysis}, volume~3.
\newblock Wiley New York, 1966.

\bibitem{feng2009combustion}
W.~D. Feng, M.~Li, M.~Li, and H.~Pu.
\newblock Combustion optimization based on rbf neural network and
  multi-objective genetic algorithms.
\newblock In {\em Genetic and Evolutionary Computing, 2009. WGEC'09. 3rd
  International Conference on}, pages 496--501. IEEE, 2009.

\bibitem{foreman2008architecture}
J.~C. Foreman.
\newblock {\em Architecture for intelligent power systems management,
  optimization, and storage}.
\newblock University of Louisville, 2008.

\bibitem{gondzio1996multiple}
J.~Gondzio.
\newblock Multiple centrality corrections in a primal-dual method for linear
  programming.
\newblock {\em Computational Optimization and Applications}, 6(2):137--156,
  1996.

\bibitem{gould2004preprocessing}
N.~Gould and P.~L. Toint.
\newblock Preprocessing for quadratic programming.
\newblock {\em Mathematical Programming}, 100(1):95--132, 2004.

\bibitem{gu2011online}
Y.~Gu, W.~Zhao, and Z.~Wu.
\newblock Online adaptive least squares support vector machine and its
  application in utility boiler combustion optimization systems.
\newblock {\em Journal of Process Control}, 21(7):1040--1048, 2011.

\bibitem{han2015deep}
S.~Han, H.~Mao, and W.~J. Dally.
\newblock Deep compression: Compressing deep neural networks with pruning,
  trained quantization and huffman coding.
\newblock {\em arXiv preprint arXiv:1510.00149}, 2015.

\bibitem{hasini2009analysis}
H.~Hasini, M.~Z. Yusoff, N.~H. Shuaib, M.~H. Boosroh, and M.~A. Haniff.
\newblock Analysis of flow and temperature distribution in a full scale utility
  boiler using cfd.
\newblock In {\em Energy and Environment, 2009. ICEE 2009. 3rd International
  Conference on}, pages 208--214. IEEE, 2009.

\bibitem{hochreiter1997long}
S.~Hochreiter and J.~Schmidhuber.
\newblock Long short-term memory.
\newblock {\em Neural computation}, 9(8):1735--1780, 1997.

\bibitem{NeuCo}
N.~Inc.
\newblock Boiler optimization software solution, 2017.

\bibitem{kusiak2005optimizing}
A.~Kusiak, A.~Burns, and F.~Milster.
\newblock Optimizing combustion efficiency of a circulating fluidized boiler: A
  data mining approach.
\newblock {\em International Journal of Knowledge-based and Intelligent
  Engineering Systems}, 9(4):263--274, 2005.

\bibitem{kusiak2006combustion}
A.~Kusiak and Z.~Song.
\newblock Combustion efficiency optimization and virtual testing: A data-mining
  approach.
\newblock {\em IEEE Transactions on Industrial Informatics}, 2(3):176--184,
  2006.

\bibitem{li2016combustion}
G.~Li and P.~Niu.
\newblock Combustion optimization of a coal-fired boiler with double linear
  fast learning network.
\newblock {\em Soft Computing}, 20(1):149--156, 2016.

\bibitem{li2017deep}
G.-Q. Li, X.-B. Qi, K.~C. Chan, and B.~Chen.
\newblock Deep bidirectional learning machine for predicting no x emissions and
  boiler efficiency from a coal-fired boiler.
\newblock {\em Energy \& Fuels}, 31(10):11471--11480, 2017.

\bibitem{liu2018pbgan}
J.~Liu, J.~Zhang, Y.~Ding, X.~Xu, M.~Jiang, and Y.~Shi.
\newblock Pbgan: Partial binarization of deconvolution based generators.
\newblock {\em arXiv preprint arXiv:1802.09153}, 2018.

\bibitem{liu2010nonlinear}
X.~Liu, P.~Guan, and C.~Chan.
\newblock Nonlinear multivariable power plant coordinate control by constrained
  predictive scheme.
\newblock {\em IEEE transactions on control systems technology},
  18(5):1116--1125, 2010.

\bibitem{mezura2006comparative}
E.~Mezura-Montes, J.~Vel{\'a}zquez-Reyes, and C.~A. Coello~Coello.
\newblock A comparative study of differential evolution variants for global
  optimization.
\newblock In {\em Proceedings of the 8th annual conference on Genetic and
  evolutionary computation}, pages 485--492. ACM, 2006.

\bibitem{nocedal2006sequential}
J.~Nocedal and S.~J. Wright.
\newblock {\em Sequential quadratic programming}.
\newblock Springer, 2006.

\bibitem{park2013numerical}
H.~Y. Park, S.~H. Baek, Y.~J. Kim, T.~H. Kim, D.~S. Kang, and D.~W. Kim.
\newblock Numerical and experimental investigations on the gas temperature
  deviation in a large scale, advanced low nox, tangentially fired pulverized
  coal boiler.
\newblock {\em Fuel}, 104:641--646, 2013.

\bibitem{pechenizkiy2010online}
M.~Pechenizkiy, J.~Bakker, I.~{\v{Z}}liobait{\.e}, A.~Ivannikov, and
  T.~K{\"a}rkk{\"a}inen.
\newblock Online mass flow prediction in cfb boilers with explicit detection of
  sudden concept drift.
\newblock {\em ACM SIGKDD Explorations Newsletter}, 11(2):109--116, 2010.

\bibitem{peng2009improved}
X.~Peng and P.~Wang.
\newblock An improved multiobjective genetic algorithm in optimization and its
  application to high efficiency and low nox emissions combustion.
\newblock In {\em Power and Energy Engineering Conference, 2009. APPEEC 2009.
  Asia-Pacific}, pages 1--4. IEEE, 2009.

\bibitem{rastegari2016xnor}
M.~Rastegari, V.~Ordonez, J.~Redmon, and A.~Farhadi.
\newblock Xnor-net: Imagenet classification using binary convolutional neural
  networks.
\newblock In {\em European Conference on Computer Vision}, pages 525--542.
  Springer, 2016.

\bibitem{snoek2012practical}
J.~Snoek, H.~Larochelle, and R.~P. Adams.
\newblock Practical bayesian optimization of machine learning algorithms.
\newblock In {\em Advances in neural information processing systems}, pages
  2951--2959, 2012.

\bibitem{szega2015optimization}
M.~Szega and G.~T. Nowak.
\newblock An optimization of redundant measurements location for thermal
  capacity of power unit steam boiler calculations using data reconciliation
  method.
\newblock {\em Energy}, 92:135--141, 2015.

\bibitem{trelea2003particle}
I.~C. Trelea.
\newblock The particle swarm optimization algorithm: convergence analysis and
  parameter selection.
\newblock {\em Information processing letters}, 85(6):317--325, 2003.

\bibitem{vapnik2013nature}
V.~Vapnik.
\newblock {\em The nature of statistical learning theory}.
\newblock Springer Science \& Business Media, 2013.

\bibitem{wang2018optimizing}
C.~Wang, Y.~Liu, S.~Zheng, and A.~Jiang.
\newblock Optimizing combustion of coal fired boilers for reducing nox emission
  using gaussian process.
\newblock {\em Energy}, 2018.

\bibitem{worldcoal}
{World Coal Association}.
\newblock {Coal and Electricity}.
\newblock Technical report, {World Coal Association}, 2016.

\bibitem{xu2013balanced}
W.~Xu and C.~Taihua.
\newblock The balanced model and optimization of nox emission and boiler
  efficiency at a coal-fired utility boiler.
\newblock In {\em Conference Anthology, IEEE}, pages 1--4. IEEE, 2013.

\bibitem{xu2018scaling}
X.~Xu, Y.~Ding, S.~X. Hu, M.~Niemier, J.~Cong, Y.~Hu, and Y.~Shi.
\newblock Scaling for edge inference of deep neural networks.
\newblock {\em Nature Electronics}, 1(4):216, 2018.

\bibitem{xu2018efficient}
X.~Xu, Q.~Lu, T.~Wang, Y.~Hu, C.~Zhuo, J.~Liu, and Y.~Shi.
\newblock Efficient hardware implementation of cellular neural networks with
  incremental quantization and early exit.
\newblock {\em ACM Journal on Emerging Technologies in Computing Systems
  (JETC)}, 14(4):48, 2018.

\bibitem{xu2018quantization}
X.~Xu, Q.~Lu, L.~Yang, S.~Hu, D.~Chen, Y.~Hu, and Y.~Shi.
\newblock Quantization of fully convolutional networks for accurate biomedical
  image segmentation.
\newblock In {\em Proceedings of the IEEE Conference on Computer Vision and
  Pattern Recognition}, pages 8300--8308, 2018.

\bibitem{xu2018resource}
X.~Xu, T.~Wang, Q.~Lu, and Y.~Shi.
\newblock Resource constrained cellular neural networks for real-time obstacle
  detection using fpgas.
\newblock In {\em 2018 19th International Symposium on Quality Electronic
  Design (ISQED)}, pages 437--440. IEEE, 2018.

\bibitem{yin2003further}
C.~Yin, L.~Rosendahl, and T.~J. Condra.
\newblock Further study of the gas temperature deviation in large-scale
  tangentially coal-fired boilers☆.
\newblock {\em Fuel}, 82(9):1127--1137, 2003.

\bibitem{zhang1985error}
G.~Zhang, R.~Veale, T.~Charlton, B.~Borchardt, and R.~Hocken.
\newblock Error compensation of coordinate measuring machines.
\newblock {\em CIRP Annals-Manufacturing Technology}, 34(1):445--448, 1985.

\bibitem{zhang2012simultaneous}
S.~Zhang, C.~W. Taft, J.~Bentsman, A.~Hussey, and B.~Petrus.
\newblock Simultaneous gains tuning in boiler/turbine pid-based controller
  clusters using iterative feedback tuning methodology.
\newblock {\em ISA transactions}, 51(5):609--621, 2012.

\bibitem{zhang2014sequential}
Y.~Zhang, H.~Dai, C.~Xu, J.~Feng, T.~Wang, J.~Bian, B.~Wang, and T.-Y. Liu.
\newblock Sequential click prediction for sponsored search with recurrent
  neural networks.
\newblock {\em arXiv preprint arXiv:1404.5772}, 2014.

\bibitem{zhang2007modeling}
Y.~Zhang, Y.~Ding, Z.~Wu, L.~Kong, and T.~Chou.
\newblock Modeling and coordinative optimization of no x emission and
  efficiency of utility boilers with neural network.
\newblock {\em Korean Journal of Chemical Engineering}, 24(6):1118--1123, 2007.

\bibitem{zhao2009modeling}
H.~Zhao and P.-h. Wang.
\newblock Modeling and optimization of efficiency and nox emission at a
  coal-fired utility boiler.
\newblock In {\em Power and Energy Engineering Conference, 2009. APPEEC 2009.
  Asia-Pacific}, pages 1--4. IEEE, 2009.

\bibitem{zhao2015fuzzy}
W.~Zhao, G.~Zhao, M.~Lv, and J.~Zhao.
\newblock Fuzzy optimization control for nox emissions from power plant boilers
  based on nonlinear optimization 1.
\newblock {\em Journal of Intelligent \& Fuzzy Systems}, 29(6):2475--2481,
  2015.

\bibitem{zheng2008combining}
L.~Zheng, H.~Zhou, C.~Wang, and K.~Cen.
\newblock Combining support vector regression and ant colony optimization to
  reduce nox emissions in coal-fired utility boilers.
\newblock {\em Energy \& Fuels}, 22(2):1034--1040, 2008.


\end{thebibliography}
\end{document}